%% file: mainv2.tex
\documentclass{article}
\usepackage[utf8]{inputenc}
\usepackage[letterpaper, margin=1in]{geometry}

\usepackage{setspace}
\usepackage{multirow}
\usepackage{graphicx}
\usepackage{dsfont}
\usepackage{amsfonts}
\usepackage{amsmath}
\usepackage{bbm,bbold}
\usepackage{float}
\usepackage{amsfonts,amsmath,amsthm,braket,framed}
 
\usepackage[normalem]{ulem}
\usepackage[dvipsnames]{xcolor}
\definecolor{DartmouthGreen}{RGB}{0, 105, 62}
\usepackage{physics}
\usepackage[utf8]{inputenc}

\makeatletter
\renewcommand*\env@matrix[1][*\c@MaxMatrixCols c]{%
  \hskip -\arraycolsep
  \let\@ifnextchar\new@ifnextchar
  \array{#1}}
\makeatother

\usepackage{graphicx}
\usepackage{subcaption}
\usepackage{mwe}

\usepackage{amsthm}

\usepackage{mathtools}
\usepackage{amssymb}
\usepackage{tikz}
\usetikzlibrary{quantikz}
\usetikzlibrary{shapes,snakes}

\usepackage{pgfplots}
\pgfplotsset{compat=1.13}

 \usepackage{graphicx}
 \usepackage[colorlinks=true, linkcolor=green, citecolor=blue]{hyperref}

 \def\be{\begin{equation}}
 \def\ee{\end{equation}}
 \def\bea{\begin{eqnarray}}
 \def\eea{\end{eqnarray}}
 \def\bean{\begin{eqnarray*}}
 \def\eean{\end{eqnarray*}}
 \def\gsim{\mathrel{\rlap{\lower0.2em\hbox{$\sim$}}\raise0.2em\hbox{$>$}}}
 \def\ksim{\mathrel{\rlap{\lower0.2em\hbox{$\sim$}}\raise0.2em\hbox{$<$}}}
 \def\kg{\mathrel{\rlap{\lower0.25em\hbox{$>$}}\raise0.25em\hbox{$<$}}}

 \newcommand\glnf[1]{$\mathrm{GL}_n\left(\mathbb{F}_2\right)$}
 
 \newtheorem{lemma}{Lemma}[section]
 
\renewcommand\vec{\mathbf}

\usepackage{enumitem}

\theoremstyle{definition}

\newtheorem{exmp}{Example}[section]
\newtheorem{remark}{Remark}[section]

\newcommand{\ketf}[1]{|{#1})}
\newcommand{\braf}[1]{({#1}|}

\usepackage{authblk}

\begin{document}
\title{A Sierpinski Triangle Fermion-to-Qubit Transform}

\author[1]{Brent Harrison}
\author[2]{Mitchell Chiew}
\author[1]{Jason Necaise}
\author[1]{Andrew Projansky}
\author[2,3,\footnote{Address from Sept 2024: Department of Computer Science, University of Oxford, Oxford OX1 3QG, UK}]{Sergii Strelchuk}
\author[1]{James D.\ Whitfield}
\affil[1]{Department of Physics and Astronomy, Dartmouth College, Hanover, New Hampshire 03755, USA}
\affil[2]{\small DAMTP, Center for Mathematical Sciences, University of Cambridge, Cambridge CB3 0WA, UK}
\affil[3]{Department of Computer Science, University of Warwick, Coventry CV4 7AL,
UK}

\maketitle
\begin{abstract}
    In order to simulate a system of fermions on a quantum computer, it is necessary to represent the fermionic states and operators on qubits. This can be accomplished in multiple ways, including the well-known Jordan-Wigner transform, as well as the parity, Bravyi-Kitaev, and ternary tree encodings. Notably, the Bravyi-Kitaev encoding can be described in terms of a classical data structure, the Fenwick tree. Here we establish a correspondence between a class of classical data structures similar to the Fenwick tree, and a class of one-to-one fermion-to-qubit transforms. We present a novel fermion-to-qubit encoding based on the recently discovered ``Sierpinski tree'' data structure, which matches the operator locality of the ternary tree encoding, and has the additional benefit of encoding the fermionic states as computational basis states. This is analogous to the formulation of the Bravyi-Kitaev encoding in terms of the Fenwick tree.
    
\end{abstract}



\section{Introduction}


Quantum simulation is one of the most compelling applications of quantum computers, with relevance to problems in quantum chemistry, condensed matter physics, and even high energy physics~\cite{georgescu_quantum_2014, kassal_simulating_2011}. Since qubits do not respect fermionic antisymmetry, any simulation of fermions on a quantum
computer first requires the construction of encoded qubit representations of fermionic states and operators. The various methods for doing so are referred to as fermion-to-qubit encodings, the most familiar of which is the Jordan-Wigner mapping~\cite{Jordan:1928wi}. This mapping encodes local fermionic operators as highly nonlocal tensor products of Pauli matrices (``Pauli strings''), which act nontrivially on $O(n)$ qubits; we say they have $O(n)$ ``Pauli weight''. 
This nonlocality significantly increases the overhead of quantum simulation, and it is desirable to construct alternatives to Jordan-Wigner that minimize the locality of the encoded operators as much as possible. 

The Bravyi-Kitaev encoding \cite{bravyi_fermionic_2002, seeley_bravyi-kitaev_2012} improves on the locality of Jordan-Wigner, reducing it from $O(n)$ to $O(\log_2 n)$. In \cite{havlicek_operator_2017}, Havlicek et al.\ show that this encoding can be represented by a classical data structure, the Fenwick tree \cite{fenwick_new_1994, fenwick_new_1996}. They also present other examples of Fenwick tree-like data structures, each of which has a corresponding fermion-to-qubit encoding. In 2019, Jiang et al.\ were able to improve this locality further, finding a provably optimal $O(\log_3 n)$ encoding based on ternary trees \cite{jiang_optimal_2020}. It was previously not clear if this encoding could also be represented by a data structure analogous to
a Fenwick tree. In this work, we develop a fermion-to-qubit transform based on a novel Fenwick tree-like data structure, which we call the Sierpinski tree. We show that this transform matches the operator locality properties of the ternary trees, while also encoding the fermionic states as computational basis states. 

The paper is organized as follows. We begin with some definitions in Sec.~\ref{sec:definitions}. In Sec.~\ref{sec:encodings} we review the Jordan-Wigner, and Bravyi-Kitaev encodings. We highlight the formulation of the latter in terms of the Fenwick tree data structure. We also review the optimal ternary tree encoding, and discuss its locality properties. In Sec.~\ref{sec:sierpinski} we introduce our novel classical data structure, the Sierpinski tree, and describe a fermion-to-qubit transform based on it. In Sec.~\ref{sec:comparison} we compare the Sierpinski tree encoding to the ternary tree encoding, and show that it has identical operator locality. In Sec.~\ref{sec:ibm} we establish a general correspondence between the class of Fenwick tree-like data structures and a class of related fermion-to-qubit transforms.  We conclude with Sec.~\ref{sec:conclusion} and present an outlook for future work.

\section{Definitions}\label{sec:definitions}
Throughout the paper we will make use of several definitions and concepts related to the Clifford group and the stabilizer formalism. The Pauli matrices are given by

\begin{equation}
I \equiv \begin{bmatrix}1 & 0 \\ 0 & 1\end{bmatrix},\ X \equiv \begin{bmatrix}0 & 1 \\ 1 & 0\end{bmatrix},\ Y \equiv \begin{bmatrix}0 & -i \\ i & 0\end{bmatrix},\ Z \equiv \begin{bmatrix}1 & 0 \\ 0 & -1\end{bmatrix}.
\end{equation}

A \textit{Pauli string} on $n$ qubits is a tensor product $q P$, where the phase $q \in K \equiv\{1, -1, i, -i\}$ and $P \in \{I,X,Y,Z\}^{\otimes n}$. The $n$-qubit \textit{Pauli group} $\mathcal{P}_n$ is the group of all such strings. We define the \textit{Pauli weight} of a Pauli string $qP$ to be the number of non-identity terms in the tensor product.

The Clifford group is the set of unitaries generated by the CNOT, Hadamard and Phase gates,

\begin{equation}
\text{CNOT} = \begin{bmatrix} 1 & 0 & 0 & 0 \\
0 & 1 & 0 & 0\\
0 & 0 & 0 & 1 \\
0 & 0 & 1 & 0\end{bmatrix}, \quad \text{H} = \frac{1}{\sqrt{2}} \begin{bmatrix}1 & 1 \\ 1 & -1\end{bmatrix}, \quad \text{P} = \begin{bmatrix}1 & 0 \\ 0 & i\end{bmatrix}\, ,
\end{equation}
i.e. 
\begin{equation}
\mathcal{C}_n \equiv \langle \mathrm{CNOT}_{ij}, \mathrm{H}_i, \mathrm{P}_i \rangle\, .
\end{equation}
The Clifford group can equivalently be defined in terms of the set of unitaries that normalizes the $n$-qubit Pauli group $\mathcal{P}_n$~\cite{gottesman_heisenberg_1998},
\begin{equation}
\mathcal{C}_n \equiv \{U \in \text{U}(2^n)\ |~ U \mathcal{P}_n U^\dagger = \mathcal{P}
_n\}\, .
\end{equation}

We will also make use of some concepts from classical computer science and graph theory. In particular, consider an $n$-bit binary array $(a_0, a_1, \dots_, a_{n-1})$, where $a_j \in \{0,1\}$. Then we can define the $k$th \emph{prefix sum} of the array as the binary sum of the first $k$ elements, $\bigoplus_{j<k}a_j$.

Let $D = \{V,E\}$ be a directed graph. Then a \emph{directed path} in $D$ is a sequence of vertices, $(v_0, v_1, \dots v_n)$ such that $(v_{i-1}, v_{i}) \in E$ for all $i \in \{1,2,\dots, n\}$.

\begin{figure}[h]
    \centering
    \includegraphics[width=0.4\linewidth]{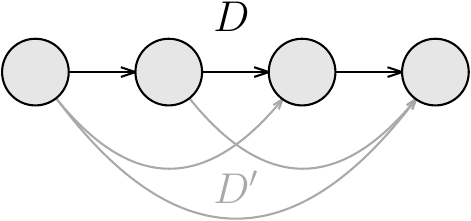}
    \caption{A directed graph $D$, which can be extended to its transitive closure $D'$ by including the gray edges.}
    \label{fig:Transitive-closurel}
\end{figure}

We say that a vertex $v_k \in V$ is \emph{reachable} from a vertex $v_j \in V$ if there exists a directed path from $v_j$ to $v_k$. Then the \emph{reachability relation} of the digraph $D$ is defined as 
\begin{equation}
R(D) = \{(v_j,v_k) \in V \times V \mid v_k \text{ is reachable from } v_j\}.
\end{equation}
We define the \emph{transitive closure} or \emph{reachability graph} of $D$ to be the graph $D' = \{V, R(D)\}$. See Fig.~\ref{fig:Transitive-closurel} for an example. We call the adjacency matrix of $D'$ the \emph{reachability matrix} of $D$.
\vspace{4mm}


Following notation in recent fermionic tensor network research \cite{obrien_local_2024, bultinck2017fermionic}, we use kets of the form $\ketf{\psi}$ to denote fermionic states, and kets of the form $\ket{\psi}$ to denote qubit states.

\section{Three Paradigmatic Encodings}\label{sec:encodings}
To put our results in perspective, in this section we will review the Jordan-Wigner, Bravyi-Kitaev and ternary tree fermion-to-qubit encodings. We will begin by discussing the formalism of second quantization for fermions.

\subsection{Fermionic States and Operators in Second Quantization}

In second quantization, the state space $\mathcal{H}_\text{fermion}$ of an $n$--mode fermionic system has a vacuum state $\ketf{\Omega_\text{vac}}$ and a \textit{Fock basis} $\{\ketf{\vec{f}} \, | \, \vec{f} \in \mathbb{F}_2^n\}$ made up of states $\ketf{\vec{f}}$ in which the occupancy of each mode is well--defined:
\begin{equation}
\begin{aligned}
	\ketf{\vec{f}} = \ketf{f_{0}, f_{1}, \dots,  f_{n-2},  f_{n-1}} \coloneqq ({a}_0^\dagger)^{f_0} ({a}_1^\dagger)^{f_1} \dots ({a}_{n-1}^\dagger)^{f_{n-1}} \ketf{\Omega_{\text{vac}}}\, . \label{eqn:fockstate}
\end{aligned}
\end{equation}
Here $a^\dagger_j$ and $a_j$ are the fermionic creation and annihilation operators for mode $j$, and $f_j \in \{0,1\}$ denotes the occupancy number of mode $j$. The state $\ketf{\Omega_\text{vac}}$ is equal to the Fock basis vector $\ketf{\vec{0}}=\ketf{0,0, \dots, 0}$ in which every fermionic mode is unoccupied; it is a simultaneous 0--eigenstate of the $n$ Hermitian number operators $\{a_j^\dagger a_j\}_{j=0}^{n-1}$, and is thus unique up to some global phase. The creation and annihilation operators respect the relations
\begin{equation}\label{eq:anticommutationRelations}
\{a_j, a_k^\dagger\} = \delta_{jk}I, \hspace{2mm} \{a_j,a_k\} = \{a_j^\dagger, a_k^\dagger\} = 0,
\end{equation}
where $\{A,B\}=AB+BA$ denotes the anti-commutator.

Due to the relations~\eqref{eq:anticommutationRelations}, the action of an $a_j^\dagger$ operator on a state $\ketf{\vec{f}} = \ketf{f_0, f_1, \dots, f_{n-1}}$ is given by

\begin{equation}\label{eq:creationOperator}
\begin{aligned}
a_j^\dagger\ketf{\vec{f}} \ \ &=  \qquad \quad a_j^\dagger \left[  (a^\dagger_0)^{f_0}(a^\dagger_1)^{n_1}\dots (a_j^\dagger)^{n_j} \quad \, \dots(a^\dagger_{n-1})^{f_{n-1}}\ketf{\vec{0}}\right] \\[0.6em]
&= \ (-1)^{p_{j}}\left[(a^\dagger_0)^{f_0}(a^\dagger_1)^{f_1}\dots (a_j^\dagger)^{f_{j+1}}\dots(a^\dagger_{n-1})^{f_{n-1}}\ketf{0,0,\dots, 0}\right]\\[0.6em]
&= \begin{cases}
    (-1)^{p_{j}}\ketf{f_0 , f_1 , \dots , f_{j-1} , 1, f_{j+1}, f_{j+2},\dots, f_{n-1}},& \text{if } f_j = 0\\
    0,              & \text{if $f_j=1$}\, ,
\end{cases}
\end{aligned}
\end{equation}
where $p_{j} \equiv {\sum_{i<j}f_i}$ is the parity of the occupied modes with index $i < j$. The factor $(-1)^{p_{j}}$ is the phase picked up by anticommuting $a_j^\dagger$ through $a_i^\dagger$ for non-zero $f_i$ with $i<j$. If $f_j=1$, then the $a_j^\dagger$ operator will meet the pre-existing $a_j^\dagger$ and annihilate the state; if $f_j=0$, then $a_j^\dagger$ instead takes up its canonical position in the definition of a new state with $f_j=1$ according to \ref{eqn:fockstate}.

Similarly,
\begin{equation}\label{eq:annihilationOperator}
a_j\ketf{\vec{f}} = \begin{cases}
    (-1)^{p_{j}}\ketf{ f_0 , f_1, f_2, \dots,  f_{j-1}, 0 , f_{j+1} , f_{j+2} , \dots f_{n-1}},& \text{if } f_j = 1\\
    0,              & \text{if $f_j=0$}\, .
\end{cases}
\end{equation}

Eqs.~\ref{eq:creationOperator} and \ref{eq:annihilationOperator} can be rewritten in terms of projection operators onto the fermionic modes as
\begin{align} 
    a_j^{\dagger} &= \sum_{\vec{f} \in \mathbb{F}_2^n\, | \, f_j = 0} (-1)^{p_j} \ketf{f_0,f_1,\dots,f_{j-1},1,f_{j+1},\dots,f_{n-1}}\braf{f_0,f_1,\dots,f_{j-1},0,f_{j+1},\dots,f_{n-1}}\\
    &= \sum_{\vec{f} \in \mathbb{F}_2^n}  (-1)^{p_j}   \bigg( \prod_{k \neq j}\mathds{1}_{k} \bigg) \ketf{1}\braf{0}_j   \ketf{\vec{f}}\braf{\vec{f}} \\
    a_j &=  \sum_{\vec{f} \in \mathbb{F}_2^n\, | \, f_j = 1} (-1)^{p_j} \ketf{f_0,f_1,\dots,f_{j-1},0,f_{j+1},\dots,f_{n-1}}\braf{f_0,f_1,\dots,f_{j-1},1,f_{j+1},\dots,f_{n-1}}\\
    &=  \sum_{\vec{f} \in \mathbb{F}_2^n} (-1)^{p_j} \bigg( \prod_{k \neq j}\mathds{1}_{k} \bigg) \ketf{0}\braf{1}_j  \ketf{\vec{f}}\braf{\vec{f}}\, ,
\end{align}
where $\mathds{1}_k = \ketf{0}\braf{0}_k + \ketf{1}\braf{1}_k$ is the fermionic identity operator on the $k$th mode. The action of the creation and annihilation operators on arbitrary Fock basis states thus amounts to
\begin{align}
    \label{eqn:fermionicoperators1}
    a_j^{\dagger} \ketf{\vec{f}} &=
        \underbrace{(-1)^{p_j}}_{{\text{parity count}}}
        \underbrace{\bigg( \prod_{k \neq j}\mathds{1}_{k} \bigg) \ketf{1}\braf{0}_j }_{\text{number update}} \ketf{\vec{f}}
        \eqqcolon (-1)^{p_j} \ketf{1}\braf{0}_j \ketf{\vec{f}}  \\
    \label{eqn:fermionicoperators2}
    a_j \ketf{\vec{f}} &= 
        \underbrace{(-1)^{p_j}}_{{\text{parity count}}}
         \underbrace{ \bigg( \prod_{k \neq j}\mathds{1}_{k} \bigg) \ketf{0}\braf{1}_j }_{{\text{number update}}}
         \ketf{\vec{f}}
        \eqqcolon (-1)^{p_j}
        \ketf{0}\braf{1}_j  \ketf{\vec{f}}
        \, ,
\end{align}
where we have omitted the identity operators from the expressions on the right-hand side.

Notice that the creation and annihilation 
 operators essentially perform two operations on Fock basis vectors, which we have annotated in Eqs.~\ref{eqn:fermionicoperators1} and \ref{eqn:fermionicoperators2}. We call these the \emph{parity count} and \emph{number update} operations; they respectively multiply the state by an overall phase $(-1)^{p_j}$, and update the fermionic occupation number $f_j$ via $\ketf{0}\braf{1}_j$ or $\ketf{1}\braf{0}_j$.
 
 In order to construct a fermion-to-qubit transform, we must find a suitable representation of the fermionic states (Eq.~\ref{eqn:fockstate}) and operators (Eqs.~\ref{eqn:fermionicoperators1} and \ref{eqn:fermionicoperators2}) on qubits. We aim to show that for an important class of fermion-to-qubit transforms, this is equivalent to defining some classical data structure that encodes an $n$-bit binary array without ancillas. The locality of the associated fermion-to-qubit encoding then scales with the time complexity of performing the classical \emph{prefix sum} and \emph{array update} operations on this encoded array.
 
 We will first illustrate this with examples, beginning with the Jordan--Wigner transform. We will then generalize in Sec.~\ref{sec:genfen}.

\subsection{Jordan-Wigner}

Suppose that we would like to represent some system of fermions in second quantization on a quantum computer. For concreteness, we will consider the fermionic state $\ketf{1,0,0,1,1}$ in which three fermions occupy modes 0, 3 and 4 of a five-mode system. The most obvious way to represent such a state on qubits is to directly map this binary string onto an identical binary string stored on a qubit register,
\begin{equation}\label{eq:JWStateMapExample}
    \ketf{\vec{f}} = \ketf{1,0,0,1,1} \rightarrow \ket{10011} \coloneqq \ket{1} \ket{0} \ket{0} \ket{1} \ket{1} = \ket{\vec{f}}\, .
\end{equation}
In general we can straightforwardly encode Fock basis states $\ketf{f_0,f_1,\dots ,f_{n-1}}$ as qubit states $\ket{f_0f_1\dots f_{n-1}}$.
It is then necessary to construct appropriate encoded qubit creation and annihilation operators that preserve the fermionic anticommutation relations~\eqref{eq:anticommutationRelations}, or, equivalently, respect Eqs.~\ref{eqn:fermionicoperators1} and \ref{eqn:fermionicoperators2} on qubits.

The Jordan--Wigner transformation~\cite{Jordan:1928wi} defines such a qubit-space representation of the operators:
\begin{equation}\label{eqn:jw}
\begin{aligned}
    a^\dagger_j \ketf{\vec{f}} =  (-1)^{p_j} \ketf{1}\braf{0}_j  \ketf{\vec{f}} & \rightarrow \prod_{k<j} Z_k \left(\frac{X_j - iY_j}{2}\right)   \ket{\vec{f}} \, ,\\
    a_j\ketf{\vec{f}} = (-1)^{p_j} \ketf{0}\braf{1}_j  \ketf{\vec{f}}&\rightarrow \prod_{k<j} Z_k \left(\frac{X_j + iY_j}{2}\right)    \ket{\vec{f}}\, ,
\end{aligned}
\end{equation}
where it is understood that these operators act as the identity on qubits with indices $i > j$.

Eq.~\ref{eqn:jw} represents a faithful qubit-system analogue of the action of the fermionic operators from Eqs.~\ref{eqn:fermionicoperators1} and \ref{eqn:fermionicoperators2}. Since $Z\ket{a} = (-1)^{a}\ket{a}$ for $a \in \{0,1\}$, we have $\prod_{k<j}Z_k \ket{\vec{f}} = (-1)^{p_j} \ket{\vec{f}}$, performing the parity count operation. The qubit operators $\frac{1}{2}(X_j \mp i Y_j)$ are equal to $\ketbra{1}{0}_j$ and $\ketbra{0}{1}_j$ respectively, and update the value of the $Z$--basis state stored in the $j$th qubit of the quantum register. This implements the number update operation.

Note that the qubit representations of the $a_j^{(\dagger)}$ in Eq.~\ref{eqn:jw} are independent of $\vec{f}$, and so an operator-specific description of the Jordan--Wigner transformation is through the representations $A_j^{(\dagger)}$ of $a_j^{(\dagger)}$ with definitions:
\begin{align}
    a_j^\dagger \rightarrow \prod_{k<j} Z_k \left(\frac{X_j - iY_j}{2} \right)   \equiv A_j^\dagger  \, , \quad a_j \rightarrow \prod_{k<j} Z_k \left(\frac{X_j + iY_j}{2} \right)  \equiv A_j\, .
\end{align}

It is often convenient to work in the basis of \textit{Majorana operators} $\{\gamma_{i}\}_{i=0}^{2n-1}$ of the fermionic algebra,
\begin{equation}
\begin{aligned}
    \gamma_{2j} \equiv a^\dagger_j + a_j\, , \quad \gamma_{2j+1} \equiv i(a^\dagger_j - a_j)\, . 
    \label{eqn:maj}
\end{aligned}
\end{equation}
From the canonical anticommutation relations  \eqref{eq:anticommutationRelations}, we find that all Majorana operators are Hermitian and mutually anticommute:
\begin{equation}\label{eq:MajoranaCommutationRelations}
    \gamma_j = \gamma_j^\dagger \, , \quad \{\gamma_{j},\gamma_k\} = 2\delta_{jk}I\, .
\end{equation}
Then, under the Jordan--Wigner transform, the Majorana operators are represented as mutually anticommuting Pauli strings $\Gamma_j \in \mathcal{P}_n$ with $O(n)$ Pauli weight:
\begin{equation} \label{eqn:jwmaj}
    \gamma_{2j} \rightarrow \left(\prod_{k<j} Z_k \right) X_j  \equiv \Gamma_{2j}\, ,\quad 
    \gamma_{2j+1} \rightarrow \left(\prod_{k<j} Z_k\right) Y_j   \equiv \Gamma_{2j+1}\, .
\end{equation}

\subsection{The Bravyi-Kitaev Encoding}

We will now discuss an alternative fermion-to-qubit encoding, the Bravyi--Kitaev transformation. This encoding improves on the locality of Jordan-Wigner, with Majorana operators of $O(\log_2 n)$ rather than $O(n)$ Pauli weight. It can be succinctly described in terms of a classical data structure, the Fenwick tree~\cite{havlicek_operator_2017}.

\subsubsection{Fenwick Trees} \label{sec:fenwicktrees}

The Fenwick tree~\cite{fenwick_new_1994, fenwick_new_1996} is a binary tree data structure often used for storing frequencies and manipulating cumulative frequency tables. The tree is used to encode binary arrays $(f_0, f_1, \dots f_{n-1})$ such that both dynamic array updates and the prefix sum operation can be accomplished in $O(\log_2 n)$ fundamental operations. 

We give pseudocode for the \textbf{fenwick} algorithm below; the instruction \textbf{fenwick}$(0,n-1)$ generates a Fenwick tree with $n$ nodes.
\begin{flalign*}
&\text{\bf{fenwick}}(S, E):&& \\
&\qquad \text{if } S \neq E:&& \\
&\qquad \qquad \text{connect } E \text{ to } \textstyle \lfloor\frac{S+E}{2}\rfloor;&& \\
&\qquad \qquad \textstyle \text{\bf fenwick} \left(S, \lfloor\frac{S+E}{2}\rfloor\right); && \\
&\qquad \qquad \text{\bf fenwick} \textstyle \left(\lfloor\frac{S+E}{2}\rfloor+1,E\right); && \\
& \qquad \text{else:} && \\
& \qquad \qquad \text{return;}
\end{flalign*}
The $n$--node Fenwick tree can be used to define a mapping from some binary array $\vec{f} = (f_0, f_1, \dots f_{n-1}) \in \mathbb{F}_2^n$ to an encoded array $\vec{q} = (q_0, q_1, \dots q_{n-1}) \in \mathbb{F}_2^n$, where the $q_i$ are defined recursively as
\begin{equation}\label{eq:recursiveFenwick}
q_i = f_i + \bigoplus_{j \in C(i)} q_j\ ,
\end{equation}
where $C(i)$ is the set of children of node $i$ in the Fenwick tree.

\begin{figure}
    \centering
    \includegraphics[width = \linewidth]{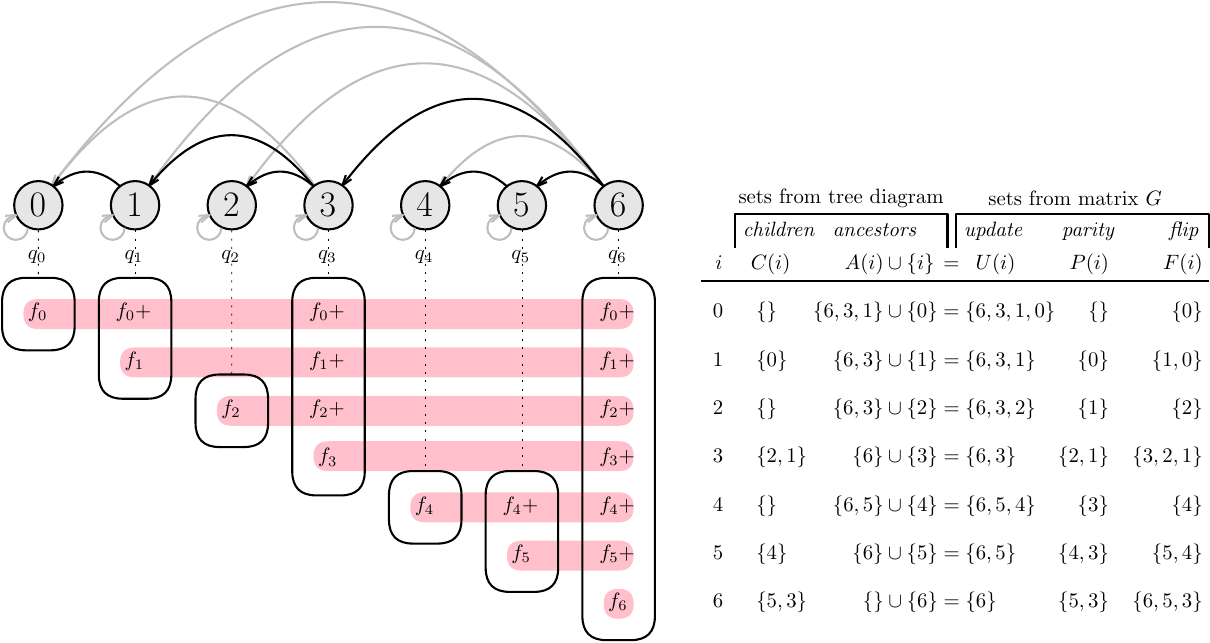}
    \caption{Fenwick tree data structure with $n=7$ nodes and a depth of 3. The boxes represent the partial sums $q_i$ of fermionic occupation numbers $f_i$ stored in each node. The edges of the Fenwick tree digraph are shown in black, while the gray arrows represent its ``completion''. The update, parity and flip sets for each node are shown on the right.}
    \label{fig:fenwickTreeSevenNodes}
\end{figure}

An equivalent way to define the encoded states from the tree is as follows. Let $R$ be the reachability matrix of the Fenwick tree, i.e.\ the adjacency matrix of the transitive closure of the tree. We then define a matrix $G = R + I$, and observe that the $q_i$ can be written as
\begin{align}
    q_i = \bigoplus_{j=0}^{n-1} G_{ij} f_j \, . \label{eq:nx}
\end{align}
We will refer to the directed graph whose adjacency matrix is $G$ as the ``completion'' of the Fenwick tree (see Fig.~\ref{fig:fenwickTreeSevenNodes}). Note that the relations \eqref{eq:nx} are invertible, and therefore $G$ is invertible over $\mathbb{F}_2$, i.e.\ $G \in \text{GL}_n(\mathbb{F}_2)$. For the example in Fig.~\ref{fig:fenwickTreeSevenNodes}, this matrix and its inverse are respectively
\begin{align}
    G = \begin{pmatrix}
        1 & 0 & 0 & 0 & 0 & 0 & 0 \\
        1 & 1 & 0 & 0 & 0 & 0 & 0 \\
        0 & 0 & 1 & 0 & 0 & 0 & 0 \\
        1 & 1 & 1 & 1 & 0 & 0 & 0 \\
        0 & 0 & 0 & 0 & 1 & 0 & 0 \\
        0 & 0 & 0 & 0 & 1 & 1 & 0 \\
        1 & 1 & 1 & 1 & 1 & 1 & 1
    \end{pmatrix} , \quad 
    G^{-1} = \begin{pmatrix}
         1 & 0 & 0 & 0 & 0 & 0 & 0 \\
         1 & 1 & 0 & 0 & 0 & 0 & 0 \\
         0 & 0 & 1 & 0 & 0 & 0 & 0 \\
         0 & 1 & 1 & 1 & 0 & 0 & 0 \\
         0 & 0 & 0 & 0 & 1 & 0 & 0 \\
         0 & 0 & 0 & 0 & 1 & 1 & 0 \\
         0 & 0 & 0 & 1 & 0 & 1 & 1
        \end{pmatrix}\, .
\end{align}
With these definitions, recall our promise that the Fenwick tree allows us to perform array update and prefix sum operations in $O(\log_2 n)$ time. This is equivalent to the claim that the size of each of the respective sets of bits we must act on to perform the update/prefix sum operations is $O(\log_2 n)$. We will now investigate these sets.


In the unencoded array $\vec{f}$, the update operation is trivially $O(1)$.  In the encoded array $\vec{q} = G \vec{f}$, updating $f_i$ requires us to update all $q_j$ that depend on $f_j$.

More formally, for each node $i$, we define an associated \emph{update set}
\begin{equation}
U(i) = \{j\in\{0,1,\dots,n-1\} \mid \text{ the partial sum } q_j \text{ contains } f_i \text{ as a term }\}
\end{equation}
This set consists precisely of the labels of node $i$ and the labels of its ancestors in the Fenwick tree, which we denote $A(i)$. We can thus express it equivalently as
\begin{itemize}
    \item $U(i) = A(i) \cup \{i\}$,
    \item $U(i) = \{ j \in \{0,1,\dots, n-1\} \, | \, G_{ji} = 1 \}$.
\end{itemize}

The size of $U(i)$  determines the time complexity of the array update operation, and 
can be shown to be $O(\log_2 n)$ for all $i$.

Determining the prefix sum $p_j$ for the unencoded array is $O(n)$. For the encoded array, let us define for each $i$ its associated \emph{parity set} $P(i)$ as the set of nodes $j$ with collective parity  $\bigoplus_{k=0}^{i-1} f_k$. We can describe $P(i)$ implicitly by the relation
\begin{equation}
\bigoplus_{k=0}^{i-1} f_k = \bigoplus_{j \in P(i)} q_i \, .
\end{equation}
We would like to define $P(i)$ explicitly, and to that end we will start by defining for each $i$ the associated \textit{flip set}, $F(i)$. This is the set of nodes $j$ whose corresponding nodes sum to $f_i$; we can find it by inverting Eq.~\ref{eq:nx}:
\begin{equation}
    f_i = \bigoplus_{j=1}^{n-1} G^{-1}_{ij} q_j =  \bigoplus_{j \in F(i)} q_j\, . \label{eqn:flipsets}
\end{equation}
Equivalently,
\begin{equation}
F(i) = \{j \in \{0,1,\dots, n-1\}\, | \, G^{-1}_{ij} = 1 \}\, .
\end{equation}

We can now define the parity set $P(i)$ in terms of the cumulative symmetric difference of the first $j{-}1$ flip sets,
\begin{equation}
P(j) = F(j-1) \, \triangle \, \dots \, \triangle \, F(1) \, \triangle \, F(0)\, ,
\end{equation}
where $A \, \triangle \, B = A \cup B \backslash (A \cap B)$ is the symmetric difference of the sets $A$ and $B$. It can be shown~\cite{seeley_bravyi-kitaev_2012, havlicek_operator_2017} that the parity sets are also of size $O(\log_2 n)$. 

\subsubsection{The Bravyi-Kitaev Encoding from the Fenwick Tree}

We can construct a fermion-to-qubit transform based on the Fenwick tree by mapping the fermionic state $\ketf{\vec{f}} = \ketf{f_0,f_1,\dots,f_{n-1}}$ to the qubit state $\ket{\vec{q}} = \ket{q_0} \ket{q_1} \dots \ket{q_{n-1}} \coloneqq \ket{q_0 q_1 \dots q_{n-1}}$. In addition to defining the encoded states, it is necessary to find the corresponding expressions for the creation and annihilation operators. 

Recall from the discussion of Eqs.~\ref{eqn:fermionicoperators1} and \ref{eqn:fermionicoperators2} that the creation and annihilation operators act as a combination of ``number update'' and ``parity count'' operations. These are respectively analogous to the array update and prefix sum operations discussed in Sec.~\ref{sec:fenwicktrees}. Indeed, we will be able to make use of the Fenwick tree update and parity sets to define the Bravyi-Kitaev creation and annihilation operators. We will first demonstrate this with an example, then generalize.

Consider the Fenwick tree--encoding for a seven--mode system. We encode the fermionic state $\ketf{f_0,f_1,\dots, f_6}$ as a qubit state $\ket{q_0 q_1 \dots q_6}$, as in Fig.~\ref{fig:fenwickTreeSevenNodes}. We then find that the fermionic creation operator $a_2^\dagger$ is mapped to the qubit creation operator $A_2^\dagger$ as follows:
\begin{equation}\label{eq:bka2}
    a_2^\dagger = (-1)^{p_2} \ketf{1}\braf{0}_2 \rightarrow  Z_1 \left(\frac{X_2 - iY_2}{2}\right) X_3 X_6 \equiv A_2^\dagger\, ,
\end{equation}
Here we have counted the parity $p_2=f_1+f_0=q_1$ of fermionic modes 0 and 1 via $Z_1$. We have also updated the occupancy $f_2 = q_2$ of the second mode by applying $\frac{1}{2}(X_2-iY_2) = \ketbra{1}{0}_2$. Finally, we have used $X_3 X_6$ to update the qubits that store the partial sums $q_3$ and $q_6$, which depend on $f_2$.

The expressions for general $A_j^\dagger, A_j$ are complicated by the fact that the number update operation acts differently on $q_j$ depending on the overall parity of its descendants in the tree. See Appendix~\ref{sec:a3} for the detailed example of $A_3^\dagger$, which is encoded as
\begin{equation}\label{eqn:a3finalmain}
A_3^\dagger = \frac{1}{2}\left(X_3 X_6 Z_1 Z_2 -i Y_3 X_6 \right) \, .
\end{equation}
As we discuss in Section \ref{sec:ibm}, Eq.~\ref{eqn:a3finalmain} is more conveniently obtained via the associated Majorana operators,
\begin{align}
    \gamma_6 &\longrightarrow  Z_1 Z_2 X_3 X_6  \, , \quad \gamma_7 \longrightarrow  Y_3 X_6 \, ,
\end{align}
and the expression $a_j^\dagger = \frac{1}{2} (\gamma_{2j} - i \gamma_{2j+1})$. In Lemma \ref{lem:upf}, we prove that the rule for obtaining the representations of the Majorana operators is
\begin{equation}
\begin{aligned}
\gamma_{2i}  \rightarrow Z_{P(i)} X_{U(i)}  \, , \quad 
\gamma_{2i+1} \rightarrow -i Z_{R(i)} X_{U(i)}  \, ,
\end{aligned} \label{eqn:ibmmaj}
\end{equation}
with the update, parity and flip sets defined, as earlier, with respect to the adjacency matrix $G \in \text{GL}_n(\mathbb{F}_2)$ of the completion of the Fenwick tree. The \textit{remainder sets} are $R(i) \equiv F(i) \, \triangle \, P(i)$. Here the subscript notation implies action on all qubits in the set, e.g. 
\begin{equation}
    Z_{P(j)} = \prod_{j \in P(j)} Z_j\, .
\end{equation}
The expressions for the creation and annihilation operators defined by the Fenwick tree encoding are thus
\begin{align}
    a_j^\dagger \longrightarrow \frac{1}{2} (Z_{P(j)} X_{U(j)}  +  Z_{R(j)} X_{U(j)} ) \equiv A_j^\dagger  \, , \quad a_j \longrightarrow \frac{1}{2} ( Z_{P(j)} X_{U(j)} - Z_{R(j)}  X_{U(j)}  ) \equiv A_j\, ,
\end{align}
which, when $n=2^k$ for some $k \in \mathbb{N}$, precisely agree with those of the Bravyi--Kitaev transformation \cite{bravyi_fermionic_2002, seeley_bravyi-kitaev_2012}, in the notation of \cite{steudtner2018fermion}.

Furthermore, in Sec.~\ref{sec:ibm} we show that Eq.~\ref{eqn:ibmmaj} gives the representations of the Majorana operators for any one-to-one fermion-to-qubit transform that encodes the fermionic states in the quantum register via $\ketf{\vec{f}} \mapsto \ket{\vec{q}} = \ket{G \vec{f}}$ for some $G \in \text{GL}_n(\mathbb{F}_2)$, not just the transform resulting from the completion of the Fenwick tree.

\subsection{Ternary Trees}

Note that our constructions of the Jordan--Wigner and Bravyi--Kitaev/Fenwick Tree encodings begin by defining a representation $\ketf{\vec{f}} \mapsto \ket{\vec{q}} = \ket{G \vec{f}}$ of second-quantized fermionic states on qubits, for some $G \in \text{GL}_n(\mathbb{F}_2)$. From this definition it is possible to derive the encoded creation/annihilation and Majorana operators. A more general approach is to instead start from a set of encoded Majorana operators and thereafter derive the states -- in which case, the transformation may not even map the $\ketf{\vec{f}}$ to computational basis states. Any set of $2n$ qubit operators that satisfy the Majorana commutation relations in Eq.~\ref{eq:MajoranaCommutationRelations} can specify a valid encoding.

In~\cite{jiang_optimal_2020}, Jiang et al.\ take this approach, introducing a fermion-to-qubit mapping that is designed to make equal use of Pauli $X$, $Y$ and $Z$ operators to construct Majoranas with provably minimal average Pauli weight, though their encoded states are in general no longer computational basis states.  Their construction is as follows.


\begin{figure} 
    \centering
    \includegraphics[width=0.7\linewidth]{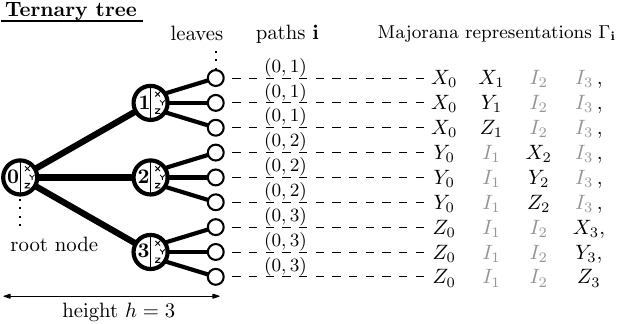}
    \caption{A four-qubit ternary tree. Each path specifies one of nine anticommuting Pauli operators, a choice of eight of them representing a set of Majorana operators.}
    \label{fig:ternaryTree}
\end{figure}

Consider a complete ternary tree of height $h$ (as in Fig 2), and associate a qubit to each node except the leaves in the rightmost level. The total number of qubits in such a tree is then given by
\begin{equation}
    n = \sum_{l = 0}^{h-1} 3^l = \frac{3^h - 1}{2}\, .
\end{equation}
Without loss of generality, we choose to label the qubits from left-to-right across the tree, starting from 0 and moving downwards in each layer of the tree. There are $3^h = 2n + 1$ unique root-to-leaf paths in the tree. Given a root-to-leaf path that traverses the qubits with labels $\vec{i} = (i_1=0,i_2,i_3,\dots,i_{h-1})$, we can write the Pauli string
\begin{equation}
    \Gamma_{\vec{i}} = (P_1)_{i_1} (P_2)_{i_2} \dots (P_{h-1})_{i_{h-1}}\, , \label{eqn:ttmaj}
\end{equation}
where $P_i$ is $X$, $Y$ or $Z$ if the root-to-leaf path takes the upper, middle or lower branch on qubit $i$, respectively, and the identity operator $I_i$ is implicit on all qubits that do not appear in the expression in Eq.~\ref{eqn:ttmaj}. These operators mutually anticommute, which we can see by comparing a given pair of operators $\Gamma_{\vec{i}}, \Gamma_{\vec{i}'}$ for $\vec{i} \neq \vec{i}'$. As the paths $\vec{i}$ and $\vec{i'}$ are distinct, they must diverge at some point. Before they diverge, the shared portion of the paths contribute to identical operators on the corresponding qubits in $\Gamma_{\vec{i}}$ and $\Gamma_{\vec{i}'}$, which commute. On the qubit of divergence, the operators $\Gamma_{\vec{i}}$ and $\Gamma_{\vec{i}'}$ act with different Pauli operators on the same qubit, which anticommute; on each qubit after the divergence, at least one of the Pauli strings will act with the identity. Thus the two operators anticommute overall, due to the single divergence point of the two paths.

Note that while there are $2n + 1$ operators $\Gamma_{\vec{i}}$, the product of these operators over all $\vec{i}$ is proportional to $I^{\otimes n}$ \cite{anticommuting2021sarkar}. Thus the total number of \emph{independent} operators in $\{\Gamma_{\vec{i}}\}_{\text{paths }\vec{i}}$ is $2n$. Any subset $\{\Gamma_{\vec{i}_k}\}_{k=0}^{2n-1}$ of $2n$ of these Pauli strings is a valid set of representations of the Majorana operators, as promised. A ternary tree transformation is thus a mapping of the form
\begin{align}
    \gamma_j \longrightarrow \Gamma_{\vec{i}_j}\, ,
\end{align}
for $j=0,1,\dots,2n-1$.

\begin{figure}
    \centering
    \includegraphics[width=0.7\linewidth]{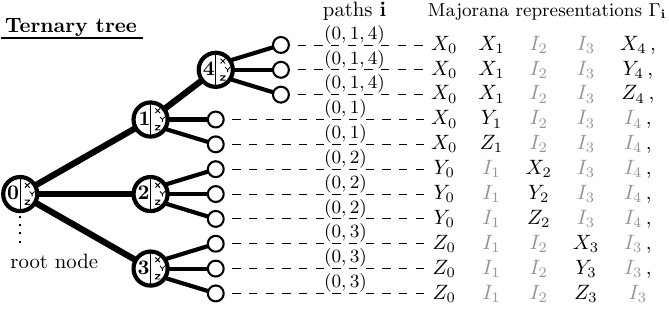}
    \caption{Any ternary tree can give rise to a set of anticommuting Pauli operators. Here a 5-qubit ternary tree produces 11 anticommuting Pauli operators.}
    \label{fig:tt2}
\end{figure}

Using the complete ternary tree as the underlying graph, this encoding leads to Majoranas with Pauli weight $\log_3(2n+1)$, which the authors of \cite{jiang_optimal_2020} prove to be optimal. The procedure that Figure \ref{fig:ternaryTree} details does not require the underlying ternary tree to be complete, and so their method extends to simulating systems of any number $n'=n+k \in \mathbb{N}$ of fermionic modes by converting $k$ of the leaves in Figure \ref{fig:ternaryTree} to labelled vertices, then repeating the procedure on this larger tree graph, as in Figure \ref{fig:tt2}. This produces $2n'+1$ anticommuting Pauli operators of weight either $\lfloor \log_3(2n'+1) \rfloor$ or $\lceil \log_3(2n'+1) \rceil$.

More generally, ternary tree transformations extend to a broader class of mappings that can involve any number of qubits. Given a possibly incomplete ternary tree with $n \in \mathbb{N}$ vertices, we can create $2n+1$ leaves by adding virtual edges so that each vertex has three children \cite{miller2023bonsai}. The end result, once again, is a set of $2n+1$ anticommuting Pauli operators, with weights depending on the structure of the the underlying tree.

\section{Generalized Fenwick Trees}\label{sec:genfen}

The Fenwick tree gives a prescription for storing a sequence of bits $(f_0,f_1,\dots, f_{n-1})$ as partial sums of those bits, $(q_0,q_1,\dots, q_{n-1})$. We can generalize this idea by considering all possible ways to encode the $f_i$ as partial sums $q_i$. We represent each of these by some invertible binary matrix $G \in \mathrm{GL}_n(\mathbb{F}_2)$, such that
\begin{equation}
q_j = \bigoplus_{i=0}^{n-1} G_{ij} f_i\, .
\end{equation}
Depending on the choice of $G$, the complexity of the array update and prefix sum $(\bigoplus_{i \leq j} f_i )$ operations will vary. In particular,
\begin{itemize}
\item A naive array implementation ($q_j = f_j$, or equivalently, $G = \mathds{1}$) leads to $O(1)$ array update complexity, and $O(n)$ prefix sum complexity. 

\item Alternatively, one could directly store the prefix sums by defining $q_j = \bigoplus_{i \leq j} f_i$. Then $G_{ij} = 1$ if $i \geq j$, $G_{ij} = 0$ otherwise. This reduces the prefix sum complexity to $O(1)$, but increases the array update complexity to $O(n)$.

\item The Fenwick tree has $O(\log_2 n)$ complexity for both operations. The corresponding $G$ matrix is given by $G = R + I$, where $R$ is the reachability matrix of the Fenwick tree digraph.
\end{itemize}

These examples are analogous to the Jordan-Wigner, Parity and Bravyi-Kitaev encodings, respectively. More generally, given a matrix $G \in \mathrm{GL}_n(\mathbb{F}_2)$, there is an associated fermion-to-qubit encoding, which can be defined by encoding the fermionic state $\ketf{\vec{f}}$ in the qubit state $\ket{G\vec{f}}$, which is a $(f_i)$--stabiliser state of the representations $A_i^\dagger A_i$ of the number operators $a_i^\dagger a_i$.




\begin{remark}
    The unitary operator that maps the computational basis state $\ket{\vec{f}}$ to $\ket{G\vec{f}}$ has a decomposition consisting entirely of CNOT gates \cite{bataille_quantum_2020}; indeed, $\mathrm{GL}_n(\mathbb{F}_2)$ is isomorphic to the subgroup of $\mathcal{U}(n)$ generated by CNOTs. One way to appreciate this is through the binary symplectic formalism (see Appendix \ref{sec:binarysymplectic}), which represents a CNOT circuit by a matrix
\begin{equation}
O = \begin{bmatrix}G & 0 \\ 0 & G^{-T}\end{bmatrix},\ G \in \mathrm{GL}_n(\mathbb{F}_2)\, .
\end{equation}

\end{remark}

\section{Sierpinski Trees}\label{sec:sierpinski}
From the discussion above, it becomes apparent that classical data structures play a central role in designing efficient fermion-to-qubit mappings. The new data structure called the Sierpinski tree\footnote{In general, the data structure will be a forest, rather than a single tree. In an abuse of nomenclature, we will nonetheless refer to it as a tree throughout.} yields a mapping which matches the operator locality of the ternary tree encoding, and has the additional benefit of encoding the fermionic states as computational basis states. This data structure is described in detail in a companion paper \cite{harrison_sierpinski_2024}, which we summarize below. It will helpful to begin by describing what we will call an ``unpruned'' Sierpinski tree. We will then use this to construct an optimized version, which we call a ``pruned'' Sierpinski tree.

\subsection{Unpruned Sierpinski Trees}

Here we present an algorithm for the construction of the unpruned tree when $n$ is a power of 3. In order to construct this tree for other values of $n$, follow the procedure to construct a tree with $3^{\lceil \log_3(n) \rceil}$ nodes, and then delete the nodes with indices $i \geq n$. The algorithm follows:
$\text{}$\newline \\
define {\bf{sierpinski}}$(S, E)$:\\
$\text{} \hspace{7mm} \text{if}$ $S \neq E$: {\color{teal} // start $\neq$ end }\\
$\text{} \hspace{14mm}$ L = $S + \frac{1}{2}(\frac{E-S+1}{3}-1)$; {\color{teal} // ``left'' point, midpoint of first third of $[S,E]$)}\\
$\text{} \hspace{14mm}$ C = $\frac{S+E}{2}$; {\color{teal} // ``center'' point, midpoint of second third of $[S,E]$}\\
$\text{} \hspace{14mm}$ R = $E - (L-S)$; {\color{teal} // ``right'' point, midpoint of final third of $[S,E]$}\\
$\text{} \hspace{14mm}$        connect $C$ to $L$;\\
$\text{} \hspace{14mm}$        connect $C$ to $R$;\\
$\text{} \hspace{14mm}$        {\color{teal}// Divide interval into thirds, apply function to each third}\\
$\text{} \hspace{14mm}$ T = $2(L-S)$;\\
$\text{} \hspace{14mm}$        {\bf sierpinski}$(S,\ S + T)$;\\
$\text{} \hspace{14mm}$        {\bf sierpinski}$(S + T + 1,\ S + 2T + 1)$;\\
$\text{} \hspace{14mm}$        {\bf sierpinski}$(S + 2T + 2,\ E))$;\\
$\text{} \hspace{7mm}$    else:\\
$\text{} \hspace{14mm}$        return;

Here the function {\bf sierpinski}$(0,3^k - 1)$ will create an unpruned Sierpinski tree with $n = 3^k$ nodes. Figs.~\ref{fig:n9} and \ref{fig:n27} show the tree for $n = 9$ and $n = 27$ respectively. 

\input{tikz_tree9}


\input{tikz_tree27}


\begin{figure}
    \centering    \includegraphics[width=0.7\linewidth]{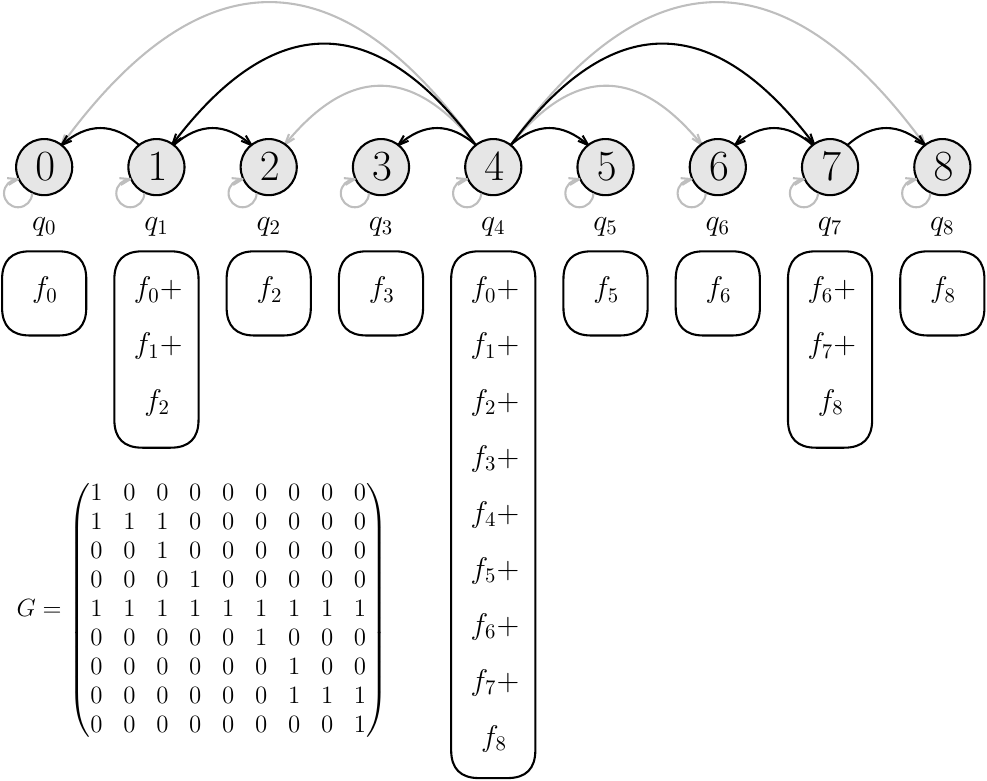}
    \caption{Diagram showing information stored in the unpruned Sierpinski tree for $n = 9$, along with the corresponding invertible binary matrix that describes its encoding of the Fock states $\ketf{\vec{f}} \mapsto \ket{\vec{q}} = \ket{G \vec{f}}$ in the computational basis.}
    \label{fig:sierpinski_nx}
\end{figure}

The tree represents the data stored analogously to the Fenwick tree, with a binary array of $n$ elements encoded in a new array of $n$ elements $q_i$, defined recursively as in Eq.~\ref{eq:nx},

\begin{equation}\label{eq:recursiveSierpinski}
q_i = f_i + \bigoplus_{j \in C(i)} q_j\ ,
\end{equation}

where $C(j)$ is the set of children of node $j$ on the tree. We illustrate this for $n = 9$ in Fig.~\ref{fig:sierpinski_nx}.

Just as with the Fenwick tree, let R be the reachability matrix of the Sierpinski tree, and let $G = R + I$. Then we can equivalently define the $q_i$ by

\begin{equation}
q_j = \bigoplus_{i=0}^{n-1} G_{ij} f_i\, .
\end{equation}

We can now define a fermion-to-qubit transform in precisely the same way as with the Fenwick tree, leading to Majoranas
\begin{equation}
\gamma_{2i} \rightarrow  Z_{P(i)} X_{U(i)} \equiv \Gamma_{2i}\, , \quad 
\gamma_{2i+1} \rightarrow -i Z_{R(i)} X_{U(i)}  \equiv \Gamma_{2i+1}\, .
\end{equation}

The Majoranas associated with node $i$ have worst-case Pauli weight $w_n(i) \equiv \abs{U(i) \cup P(i)} + 1$.  In our companion paper \cite{harrison_sierpinski_2024}, we prove that this quantity satisfies
\begin{equation}
w_n(j) \leq \lceil \log_3 n \rceil + 1
\end{equation}
In \cite{jiang_optimal_2020}, Jiang et al.\ prove that that the average Pauli weight $\overline{w}$ of a one-to-one fermion-to-qubit encoding must satisfy
\begin{equation}\label{eq:lowerBound}
\overline{w} \geq \log_3 (2n) = \log_3(n) + \log_3 2 \approx \log_3n + 0.63, 
\end{equation}
so the encoding defined by the full Sierpinski tree is already close to optimal. We can nonetheless improve it further.

\subsection{Pruning the tree}

We can use a simple greedy algorithm to improve the tree. Starting with a full Sierpinski tree, for each edge in turn, delete the edge if doing so would improve the average Pauli weight of the encoding. If necessary, repeat this process until it converges. A ``pruned'' Sierpinski tree with 18 nodes is shown in Fig.~\ref{fig:prunedTree}.

\begin{figure}[h]
\centering
\scalebox{0.75}{
  \begin{tikzpicture}
      \draw
        (0, 0) node[circle,draw=black,fill=black!4,minimum size=1.05cm, label=center:\Large 0](0){}
        (1, 2) node[circle,draw=black,fill=black!4,minimum size=1.05cm, label=center:\Large 1](1){}
        (2, 0) node[circle,draw=black,fill=black!4,minimum size=1.05cm, label=center:\Large 2](2){}
        (2, 3) node[circle,draw=black,fill=black!4,minimum size=1.05cm, label=center:\Large 3](3){}
        (3, 5) node[circle,draw=black,fill=black!4,minimum size=1.05cm, label=center:\Large 4](4){}
        (4, 3) node[circle,draw=black,fill=black!4,minimum size=1.05cm, label=center:\Large 5](5){}
        (4, 0) node[circle,draw=black,fill=black!4,minimum size=1.05cm, label=center:\Large 6](6){}
        (5, 2) node[circle,draw=black,fill=black!4,minimum size=1.05cm, label=center:\Large 7](7){}
        (6, 0) node[circle,draw=black,fill=black!4,minimum size=1.05cm, label=center:\Large 8](8){}
        (4, 7) node[circle,draw=black,fill=black!4,minimum size=1.05cm, label=center:\Large 9](9){}
        (5, 9) node[circle,draw=black,fill=black!4,minimum size=1.05cm, label=center:\Large 10](10){}
        (6, 7) node[circle,draw=black,fill=black!4,minimum size=1.05cm, label=center:\Large 11](11){}
        (6, 10) node[circle,draw=black,fill=black!4,minimum size=1.05cm, label=center:\Large 12](12){}
        (7, 12) node[circle,draw=black,fill=black!4,minimum size=1.05cm, label=center:\Large 13](13){}
        (8, 10) node[circle,draw=black,fill=black!4,minimum size=1.05cm, label=center:\Large 14](14){}
        (8, 7) node[circle,draw=black,fill=black!4,minimum size=1.05cm, label=center:\Large 15](15){}
        (9, 9) node[circle,draw=black,fill=black!4,minimum size=1.05cm, label=center:\Large 16](16){}
        (10, 7) node[circle,draw=black,fill=black!4,minimum size=1.05cm, label=center:\Large 17](17){};
      \begin{scope}[->]
        \draw (1) to (0);
        \draw (1) to (2);
        \draw (4) to[bend right](1);
        \draw (4) to[bend left](7);
        \draw (4) to (3);
        \draw (4) to (5);
        \draw (7) to (6);
        \draw (7) to (8);
        \draw (10) to (9);
        \draw (10) to (11);
        \draw (13) to[bend right](10);
        \draw (13) to (12);
        \draw (13) to (14);
        \draw (16) to (15);
        \draw (16) to (17);
      \end{scope}
    \end{tikzpicture}}
    \caption{The pruned Sierpinski tree for $n = 18$.}
    \label{fig:prunedTree}
\end{figure}

\begin{figure}[btp]
    \centering
    \includegraphics{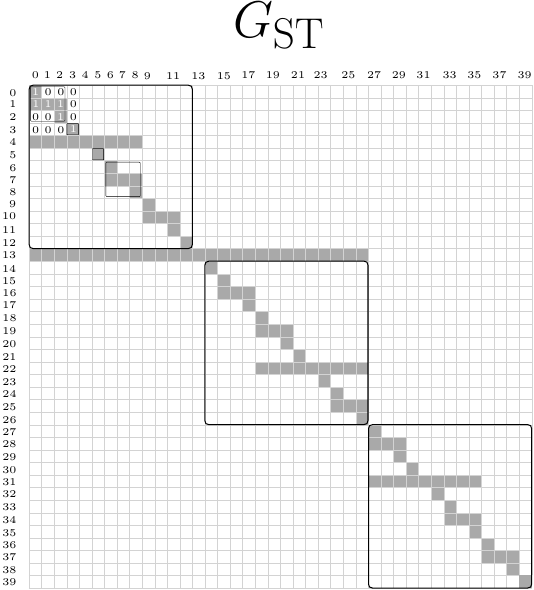}
    \caption{Invertible binary matrix corresponding to the pruned Sierpinski tree transformation for $n=40$.}
    \label{fig:ttmat}
\end{figure}

\section{Comparison to Optimal Ternary Trees}\label{sec:comparison}

From the construction of the ternary tree, we note that the average Pauli weight of its Majoranas may be obtained as follows.

If $2n+1$ is a power of three i.e. $k = \log_3(2n+1)$ is an integer, then all $2n+1$ anticommuting Pauli strings have Pauli weight equal to $k$. Now consider increasing $n$ by one. This leads to replacing one Pauli string of weight $k$ with three Pauli strings of weight $k+1$. This gives us a recurrence relation for the \emph{total} Pauli weight $T_n$ of the $2n+1$ strings,
\begin{equation}
    T_n = T_{n-1} + 3(k+1) - k = T_{n-1} + 2k + 3,
\end{equation}
which is readily solved for the average Pauli weight, 
\begin{equation}\label{eq:piecewise}
    \overline{w}_n = \begin{cases}
         \frac{5n-2}{2n+1} \qquad \qquad  & 1 \leq n < 4 \\
         \frac{7n -10}{2n+1} \qquad \qquad & 4 \leq n < 13 \\
         \quad \vdots \qquad & \quad \vdots \\
         \frac{(2k+3)n + k-\frac{3}{2}(3^k-1)}{2n+1}  \qquad & \frac{1}{2}(3^{k}-1) \leq n < \frac{1}{2}\left(3^{k+1}-1\right) 
    \end{cases}.
\end{equation}

Numerically comparing the average Pauli weight of the pruned Sierpinski tree encoding against this expression, we observe that they perfectly coincide up to large values of $n$.

\section{An algebraic view on fermion--qubit mappings that preserve the computational basis} \label{sec:ibm}
 
In this section, we explore how Clifford operators link the operator--based and state--based descriptions of fermion-to-qubit mappings.
In the process, we derive the formula for the Majorana operators of mappings that perform the encoding $\ketf{\vec{f}} \mapsto \ket{\vec{q}} = \ket{G\vec{f}}$ of the Fock basis states for some invertible binary matrix $G \in \text{GL}_n(\mathbb{F}_2)$, proving the algebraic description that appears at the end of Section~\ref{sec:genfen}.

\subsection{Fermion-to-qubit mappings defined by Clifford operators}

As we have observed, a fermion-to-qubit encoding can be thought of equivalently as a mapping of the fermionic Majorana operators to qubit Majorana operators, or as a mapping of the fermionic Fock states to qubit states. A third perspective is that each one-to-one encoding can be associated with some unitary $U$ that maps the Jordan-Wigner basis states $\ket{\vec{f}}$ and  Majoranas $\Gamma_i$ to those of the new encoding,
\begin{equation}
\ket{\vec{f}} \mapsto U\ket{\vec{f}}, \quad 
\Gamma_i \mapsto U \Gamma_i U^\dagger\, .
\end{equation}
Recall from Sections \ref{sec:encodings} and \ref{sec:sierpinski} that a fermion--to--qubit mapping arising from the generalised Fenwick tree construction corresponds to a Fock basis encoding of $\ketf{\vec{f}}$ as the qubit state $\ket{G\vec{f}}$ for some $G \in \text{GL}_n(\mathbb{F}_2)$. We are therefore interested in the class of unitaries $C_{G} \in U(2^n)$ such that $C_G \ket{\vec{f}} = \ket{G \vec{f}}$. Since $C_G$ is a linear transformation of the computational basis vectors in $\mathbb{F}_2^n$, it is a Clifford transformation \cite{dehaene2003clifford}. In fact, it is necessarily a circuit of CNOTs~\cite{bataille_quantum_2020}, and has the symplectic representation
	\begin{align}
		[C_G] = \left[\begin{array}{ccc|ccc}
			&   &   &  0  & \dots     & 0  \\ 
			&G & & \vdots & \ddots & \vdots \\ 
			&    &    &  0 & \dots     & 0 \\
			\hline
			0&\dots &0 & & & \\ 
			\vdots& \ddots & \vdots  &\multicolumn{3}{c}{(G^{-1})^\top}  \\ 
			0& \dots  &0 & & &
		\end{array}\right]\, , \label{eqn:tableau}
	\end{align}
where the $i$th column of $[C_G]$ is equal to the symplectic representation $\phi(C_G X_i C_G^\dagger)$ for $0 \leq i <n$, and for $n \leq i < 2n$ the $i$th column of $[C_G]$ equals $\phi(C_G Z_i C_G^\dagger)$, using the notation of Appendix \ref{sec:binarysymplectic}.



\begin{figure}[btp]
		\centering
		\includegraphics[width=0.9\linewidth]{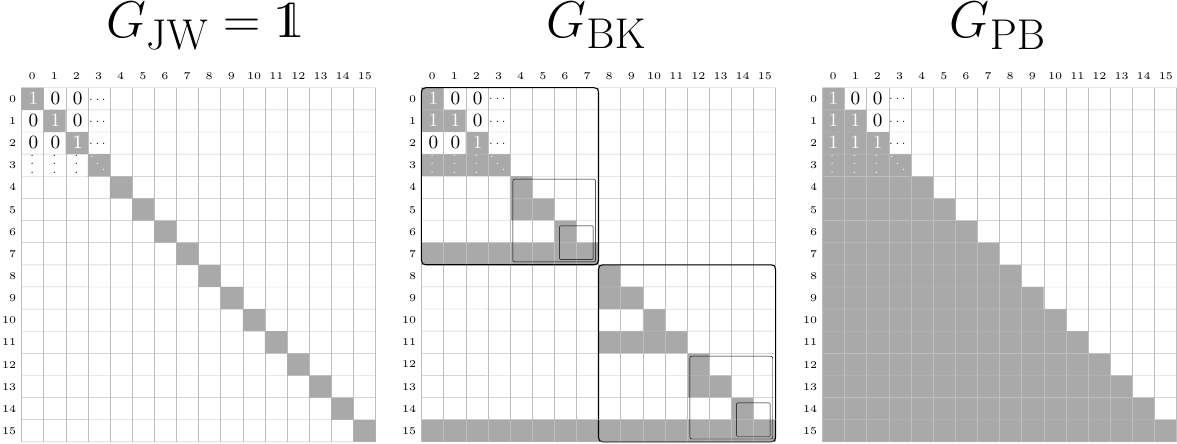}
		\caption{The invertible binary matrices $G_\text{JW}=\mathds{1}$, $G_\text{BK}$ and $G_\text{PB}$ that generate the Jordan--Wigner, Bravyi--Kitaev and parity basis transformations, respectively, for $n=16$. Shaded squares indicate entries that are equal to 1.}
		\label{fig:bkmat}
	\end{figure}

	
\begin{exmp} \textit{(State--based perspective of well-known fermion--qubit mappings with encodings $\ketf{\vec{f}} \mapsto \ket{G \vec{f}}$ for some invertible binary matrix $G$.)}
	\begin{enumerate}[label=\alph*)]
		\item The Jordan--Wigner transformation is the encoding $\ketf{\vec{f}} \mapsto \ket{\vec{f}}$ and thus has $G = \mathbb{1}_{n}$ and $C_{G} = \mathds{1}_{2^n}$.
		\item If $n$ is a power of 2, the Bravyi--Kitaev transformation is the encoding $\ketf{\vec{f}} \mapsto \ket{G\vec{f}}$, where $G=G_\text{BK}$ is recursively--defined; Figure \ref{fig:bkmat} contains the $n=16$ case.
		\item The parity basis transformation is the encoding $\ketf{\vec{f}} \mapsto \ket{G\vec{f}}$ where $G = G_\text{PB}$ is the lower triangular matrix in $\text{GL}_n(\mathbb{F}_2)$. The Pauli representations of the Majorana operators for the parity basis transformation are \cite{seeley_bravyi-kitaev_2012, miller2023bonsai}
        \begin{align}
        \gamma_{2i} \mapsto Z_{i-1}\left(\prod_{k=i+1}^{n-1} X_k \right) \, , \quad \gamma_{2i+1} \mapsto Y_i \left(\prod_{k=i+1}^{n-1} X_k \right)  \, .
        \end{align}
			
		\end{enumerate}
	\end{exmp}
	
Recall the update, flip, parity and remainder sets of an invertible binary matrix $G$ as discussed in Section \ref{sec:fenwicktrees}. Purely in terms of $G$, their definitions are as follows \cite{steudtner2018fermion}:
		\begin{itemize}
			\item The \textit{update set of $i$} is $U(i) = \{j \in \{0,1,\dots,n-1\} \, | \, G_{ji}= 1 \}$; i.e.\ the set $U(i)$ contains the indices of the rows of $G$ that have non-zero elements in column $i$.
			\item The \textit{flip set of $i$} is $F(i) = \{j \in \{0,1,\dots,n-1\}\, | \, (G^{-1})_{ij} = 1\}$; i.e.\ the set $F(i)$ contains the indices of the columns of $G^{-1}$ that have non-zero elements in row $i$.
			\item The \textit{parity set of $i$} is $P(i) = \{ j \in \{0,1,\dots,n-1\}\, | \, (\Pi G^{-1})_{ij} = 1 \}$, where $\Pi$ is the lower-triangular matrix of all 1s:
			\begin{align}
				\Pi = \begin{pmatrix}
					0 & 0 & 0 & \dots & 0 & 0 \\
					1 & 0 & 0 & \dots & 0 &0 \\
					\vdots & \vdots & \vdots & \ddots & \vdots & \vdots \\
					1 & 1& 1 & \dots & 0 &0 \\
					1 & 1& 1 & \dots & 1 & 0
				\end{pmatrix} \in \text{M}_n(\mathbb{F}_2)\, ;
			\end{align}
			i.e.\ the set $P(i)$ contains the index of each column of $G^{-1}$ for which the sum of the first $(i{-}1)$ elements of that column is nonzero modulo 2.
			\item The \textit{remainder set of $i$} is $R(i) = F(i) \, \triangle \, P(i)$, i.e.\ the set $R(i)$ contains all the elements that are not common to both $P(i)$ and $F(i)$.
		\end{itemize}
		
\begin{lemma} \label{lem:upf}
	\textit{(Determining the Clifford  $C_G$ with $C_G \ket{\vec{f}} = \ket{G\vec{f}}$ and the Pauli representations of the Majorana operators in the fermion--qubit mapping with encoding $\ketf{\vec{f}} \mapsto \ket{G\vec{f}}$.)} \\ 
	Let $G \in \emph{GL}_n(\mathbb{F}_2)$ be an invertible binary matrix. Then there exists a unique Clifford $C_G \in \mathcal{C}_n$ such that $C_G \ket{\vec{f}} = \ket{G \vec{f}}$ for all $\vec{f} \in \mathbb{F}_2^n$. The Pauli representations of the Majorana operators in the fermion--qubit mapping that encodes the Fock basis state $\ketf{\vec{f}}$ in the qubit state $\ket{G \vec{f}}$ for all $\vec{f} \in \mathbb{F}_2^n$ are:
	\begin{alignat}{2}
		\gamma_{2i} &\mapsto C_G \Gamma_{2i} C_G ^\dagger  &&= Z_{P(i)} X_{U(i)} \, , \label{eqn:upf1} \\
		\gamma_{2i+1} &\mapsto C_{G} \Gamma_{2i+1} C_{G}^\dagger && = -i Z_{R(i)} X_{U(i)} \, , \label{eqn:upf2}
	\end{alignat}
    where $\{\Gamma_{i}\}_{i=0}^{2n-1}$ are the Jordan--Wigner representations of the Majorana operators from Eq.~\ref{eqn:maj}.
		
	\begin{proof}
    From Eq.~\ref{eqn:tableau}, it follows that
	\begin{align}
    	C_G Z_i C_G^\dagger  = Z_{F(i)}\, , \quad C_G X_i   C_G^\dagger = X_{U(i)}\, , \quad C_G Y_i C_G^\dagger = -i (C_G Z_i C_G^\dagger)(C_G X_i C_G^\dagger) =  -iZ_{F(i)} X_{U(i)}\, .
	\end{align}
	
	For Eq.~\ref{eqn:upf1}, observe:
		\begin{align}
			\gamma_{2i} \mapsto C_G \Gamma_{2i} C_G^\dagger  = C_G(Z_0 Z_1 \dots Z_{i-1} X_i) &= C_G(Z_0) C_G(Z_1) \dots C_G(Z_{i-1}) C_G(X_i) \\
			&= Z_{F(0)}Z_{F(1)} \dots Z_{F(i-1)} X_{U(i)} \\
			&= Z_{P(i)} X_{U(i)} \, .\label{eqn:Xup}\\
		\end{align}
        Similarly, since $Y_i = -iZ_i X_i$, for Eq.~\ref{eqn:upf2} observe
		\begin{align}
			\gamma_{2i+1} \mapsto C_G \Gamma_{2i+1} C_G^\dagger  = -i C_G(Z_0 Z_1 \dots Z_{i-1} Z_i X_i) &= - iC_G(Z_0) C_G(Z_1) \dots C_G(Z_{i-1}) C_G(Z_i) C_G(X_i) \\
				&= -iZ_{F(0)}Z_{F(1)} \dots Z_{F(i-1)} Z_{F(i)}X_{U(i)} \\
				&= -iZ_{P(i)} Z_{F(i)} X_{U(i)}\\
				&= -iZ_{P(i) \triangle F(i)} X_{U(i)} \\
				&= -iZ_{R(i)} X_{U(i)} \label{eqn:Yup} \, .\\
			\end{align}	
        This proves the claim.
			
			\textit{Uniqueness:} The symplectic matrix in Eq.~\ref{eqn:tableau} actually identifies a family of Cliffords $\{C = C_Ge^{i\theta} \, | \,\theta \in [0,2\pi)\}$, which are the \textit{only} Cliffords to satisfy Eqs.~\ref{eqn:upf1} and \ref{eqn:upf2}. These are the operators that satisfy $C\ket{\vec{f}} = e^{i\theta} \ket{G \vec{f}}$ for $\theta \in [0,2\pi)$; we identify $C_G$ with the unique $\theta=0$ representative of this family.
		\end{proof}
	\end{lemma}
	
	\begin{lemma}\label{lem:upf2}
		For any $G \in \emph{GL}_n(\mathbb{F}_2)$, the update, parity, flip and remainder sets of $G$ satisfy
		\begin{enumerate}[label=\alph*)]
			\item $|U(i) \cap F(i) | $ is odd for all $i \in \{0,1,\dots, n-1\}$,  \label{lem:upfa}
			\item $|U(i) \cap P(i) |$ is even for all $i \in \{0,1,\dots,n-1\}$, and \label{lem:upfb}
			\item $|U(i) \cap R(i) |$ is odd for all $i \in \{0,1,\dots,n-1\}$. \label{lem:upfc}
		\end{enumerate}
		\begin{proof}
			The relevant fact is that the rows of $G^{-1}$ and the columns of $G$ are orthonormal.  Because $G^{-1} G=\mathds{1}$, the dot product of the $i$th row of $G^{-1}$ and the $i$th column of $G$ must be 1 (mod 2). Thus, there must be an odd number of shared elements of $F(i)$ and $U(i)$ for each $i \in \{0,1,\dots,n-1\}$, proving \ref{lem:upfa}. Part \ref{lem:upfb} follows from the orthonormality of the rows of $G^{-1}$ and the columns of $G$: since the $i$th row of $\Pi G^{-1}$ is equal to the sum of the first $(i-1)$ rows of $G^{-1}$, each of which is orthogonal to the $i$th column of $G$, the $i$th row of $\Pi G^{-1}$ must itself be orthogonal to the $i$th column of $G$. Therefore there must be an even number of shared elements of $P(i)$ and $U(i)$. Part \ref{lem:upfc} follows from parts \ref{lem:upfa} and \ref{lem:upfb}, and the definition $R(i) = F(i) \, \triangle \, P(i)$.
		\end{proof}
	\end{lemma}



\section{Conclusion}\label{sec:conclusion}

In summary, we have presented a novel fermion-to-qubit encoding based on the Sierpinski tree data structure. This encoding reproduces the optimal Pauli weight of the ternary tree encoding, which also representing the fermionic states as computational basis states. 

We have discussed multiple equivalent ways to think about one-to-one fermion-to-qubit encodings: as encoded representations of Fock states, as encoded representations of Majorana operators, and in terms of unitary operators which act on the Jordan-Wigner states and operators. Using this third picture, we have established a correspondence between a class of classical data structures similar to the Fenwick tree, and a corresponding class of fermion-to-qubit encodings.

Future work might include variants of these encodings tailored to particular hardware constraints, such as qubit connectivity, as well as the form of the fermionic Hamiltonian. The classical Sierpinski tree data structure was inspired by work on fermion-to-qubit transforms, and it would also be of interest to investigate other ``quantum-inspired'' classical algorithms.

\section{Acknowledgements}

The authors thank Peter Winkler, Ojas Parekh, and Joseph Gibson for helpful discussions. The authors were supported by the US NSF grant PHYS-1820747. Additional support came from the NSF (EPSCoR-1921199) and from the Office of Science, Office of Advanced Scientific Computing Research under program Fundamental Algorithmic Research for Quantum Computing. This paper was also supported by the ``Quantum Chemistry for Quantum Computers'' project sponsored by the DOE, Award DE- SC0019374. MC received the support of a Cambridge Australia Allen \& DAMTP Scholarship. JDW holds concurrent appointments at Dartmouth College and as an Amazon Visiting Academic. This paper describes work performed at Dartmouth College and is not associated with Amazon.
SS acknowledges support from the Royal Society University Research Fellowship and ``Quantum simulation algorithms for quantum chromodynamics" grant (ST/W006251/1) and EPSRC Reliable and Robust Quantum Computing grant
(EP/W032635/1).

\bibliographystyle{unsrt}
\bibliography{citations}

\appendix

\section{The Binary Symplectic Formalism}\label{sec:binarysymplectic}

In this appendix we briefly review the binary symplectic representation of Pauli strings and Cliffords.
\subsection{Binary Symplectic Representation of Pauli Strings}

Disregarding phases, the Pauli matrices can be represented by binary symplectic vectors,
\begin{equation}
I \rightarrow \begin{bmatrix}1 \\ 0\end{bmatrix},\ X \rightarrow \begin{bmatrix}1 \\ 0\end{bmatrix},\ Y \rightarrow \begin{bmatrix}1 \\ 1\end{bmatrix},\ 
Z \rightarrow \begin{bmatrix}0 \\ 1\end{bmatrix}.
\end{equation}
More generally, we can write a homomorphic map $\phi$ sending a Pauli string $p \in \mathcal{P}_n$ to a binary vector $\phi(p) \in \mathbb{F}_2^{2n}$ \cite{gunderman_transforming_2023},
\begin{equation}
\phi\left(q\prod_{k=1}^n X_k^{a_k}Z_k^{b_k}\right) = \left(\bigoplus_{k=1}^n \begin{bmatrix}a_k\end{bmatrix}\right)\bigoplus\left(\bigoplus_{k=1}^n \begin{bmatrix}b_k\end{bmatrix}\right),
\end{equation}
where $a_k, b_k \in \{0,1\}$, and $q \in \{1,-1,i,-i\}$. Then, for example,
\begin{equation}
\phi\left(X_1Y_2Z_3\right) = \phi(X_1X_2Z_2Z_3) = \begin{bmatrix} 1 \\ 1 \\ 0 \end{bmatrix} \bigoplus \begin{bmatrix} 0 \\ 1 \\ 1 \end{bmatrix} =   \begin{bmatrix}
    1 \\ 1 \\ 0 \\ 0 \\ 1 \\ 1
\end{bmatrix}.
\end{equation}
Note that the multiplication of Paulis becomes binary addition in the symplectic formalism,
\begin{equation}
\phi(p_1 p_2) = \phi(p_1) \oplus \phi(p_2).
\end{equation}

\subsection{Binary Symplectic Representation of Clifford Operators}

A Clifford tableau  \cite{aaronson_improved_2004} is a unique representation of a Clifford operation $C \in \mathcal{C}_n$, which is fully specified given the action of $C$ on a set of generators of the Pauli group. We can represent such a tableau by a matrix whose columns are the binary symplectic vectors corresponding to the Pauli strings
\begin{equation}\label{eq:cxc_czc}
C X_k C^\dagger,\ C Z_k C^\dagger,\hspace{2mm}  k \in \{1,2,\dots, n\}
\end{equation}
where the sets $\{X_k\}$ and $\{Z_k\}$ together generate the full Pauli group. Note that despite the fact that we have quotiented out global phases in our definition of $\mathcal{C}_n$, the expressions \eqref{eq:cxc_czc} are signed, with different choices of signs corresponding to different Cliffords. An additional row can be added to the tableau to represent these signs, with $+$ represented by 1 and $-$ by 0.

Then, for example, we can represent the CNOT gate by the tableau

\begin{equation}
\mathrm{CNOT} \rightarrow \left [  
\begin{array}{cccc}
    1 & 0 & 0 & 0 \\
    1 & 1 & 0 & 0 \\
    0 & 0 & 1 & 1 \\
    0 & 0 & 0 & 1 \\ \hline
    1 & 1 & 1 & 1
\end{array}
\right ]\, .
\end{equation}

These tableaux are equivalent up to signs to the symplectic group 
\begin{equation}
\mathrm{Sp}[2n, \mathbb{F}_2] \equiv \{M \in \mathrm{GL}_{2n}(\mathbb{F}_2)\mid M^T \Omega_n M = \Omega_n\},\hspace{2mm} \Omega_n = \begin{bmatrix}
  0 & I_n\\
  I_n & 0
\end{bmatrix}.
\end{equation}

\section{Derivation of Eq.~\ref{eqn:a3finalmain}}\label{sec:a3}

Here we derive the expression \eqref{eqn:a3finalmain} for $A_3^\dagger$, encoded per the 7-qubit Bravyi-Kitaev transformation.


The example of $a_2^\dagger$ \eqref{eq:bka2} is straightforward because in this case the qubit with label 2 stores only the occupation number of the second fermionic mode, so $\ketf{1}\braf{0}_2 \rightarrow \ketbra{1}{0}_2 X_{A(2)} = \frac{1}{2}(X_2-iY_2) X_3 X_6$, which is a product of local operations. An example of the case where this is not true is $f_3 = q_0+q_1+q_2$. From first principles, the process to determine the encoding $A_3^\dagger$ of $a_3^\dagger$ is as follows. Consider the action of $a_3^\dagger$ on a Fock basis vector:
\begin{align}
    a_3^\dagger \ketf{\vec{f}} &= (-1)^{p_3} \ketf{1}\braf{0}_3  \ketf{\vec{f}} = (-1)^{f_0+f_1+f_2} \bigg(\prod_{k \neq 3} \mathds{1}_k \bigg) \ketf{1}\braf{0}_3 \ketf{\vec{f}} \, . \label{eqn:a3before}
\end{align}
Rewriting the fermionic indices $f_i$ in terms of the the Fenwick tree encoding $q_i$, the fermionic state and parity count operators in the right-hand-side of Eq.~\ref{eqn:a3before} are 
\begin{align}
    \ketf{\vec{f}} = \ketf{f_0,f_1,f_2,f_3,f_4,f_5,f_6} &= \ketf{q_0, \, q_0{+}q_1, \, q_2, \, q_1 {+} q_2 {+} q_3, \, q_4, \, q_4{+}q_5, \, q_3{+}q_5{+}q_6} \\
    (-1)^{f_0 + f_1 + f_2} &= (-1)^{q_1+q_2}\, ,
\end{align}
and the number update operator is the following sum of fermionic projectors
\begin{align} \label{eqn:a3full}
    \bigg( \prod_{k \neq 3} \mathds{1}_k \bigg) \ketf{1}\braf{0}_3 = \sum_{\mathclap{\substack{q_i \in \{0,1\} \\ q_1{+}q_2{+}{q_3}=0}}} \ketf{q_0, \, q_0{+}q_1, \, q_2, \, 1, \, q_4, \, q_4{+}q_5, \, q_3{+}q_5{+}q_6}
                \braf{q_0, \, q_0{+}q_1, \, q_2, \, 0, \, q_4, \, q_4{+}q_5, \, q_3{+}q_5{+}q_6}\, .
\end{align}
The resulting state is
\begin{align} \label{eqn:cases1}
    a_3^\dagger \ketf{\vec{f}}    &= \begin{cases}
        (-1)^{q_1+q_2} \ketf{q_0', q_0'{+}q_1', q_2', 1, q_4', q_4'{+}q_5', q_3'{+}q_5'{+}q_6'}\, , & q_1+q_2+q_3 = 0 \, ,\\
        0\, , & q_1 + q_2 + q_3 = 1 \, ,
    \end{cases}
\end{align}
for some $q_i' \in \{0,1\}$ which we will now evaluate. Note that the fermionic projector in Eq.~\ref{eqn:a3full} fixes the values of the components $q_0$, $q_2$ and $q_4$, and hence it also fixes the values of $q_1$ and $q_5$. That is, $q_i' = q_i$ for each of these variables. The projector has no support on fermionic states with $q_1+q_2+q_3=1$. Suppose that $q_1+q_2+q_3=0$: the projector flips the value of $q_3$ from $0 \rightarrow 1$ if $q_1+q_2=0$, or it flips $q_3$ from $1 \rightarrow 0$ if $q_1+q_2=1$. In either case, the value of $q_3$ changes, and so $q_3' = \overline{q_3}$. Finally, the projector fixes $q_3+q_5+q_6$ and so we require $q_3'+q_5+q_6' = \overline{q_3} + q_5 + q_6' \implies q_6' = \overline{q_6}$. Therefore, Eq.~\ref{eqn:cases1} is
\begin{align}
    a_3^\dagger \ketf{\vec{f}}
    &= \begin{cases}
        (-1)^{q_1+q_2} \ketf{q_0, q_0{+}q_1, q_2, 1, q_4, q_4{+}q_5, \overline{q_3}{+}q_5{+}\overline{q_6}}\, , & q_1+q_2+q_3 = 0 \, ,\\
        0\, , & q_1 + q_2 + q_3 = 1 \, .
    \end{cases}
\end{align}

Thus, upon an array of qubits containing the Fenwick tree encoding of the fermionic occupation number states, the qubit representation $A_3^\dagger$ of the operator $a_3^\dagger$ performs the operation
\begin{align}
    A_3^\dagger :\ket{q_0 q_1q_2q_3q_4q_5q_6} \mapsto (-1)^{q_1+q_2} \ket{q_0q_1q_2\overline{q_3} q_4 q_5 \overline{q_6}}\, .
\end{align}
To represent this transformation on a qubit register storing $\ket{\vec{q}} = \ket{q_0q_1q_2q_3q_4q_5q_6}$, we require
\begin{align}
A_3^\dagger &= \underbrace{\bigg( \big( \ketbra{00}{00}_{12} + \ketbra{11}{11}_{12} \big) \ketbra{1}{0}_3 + \big( \ketbra{01}{01}_{12} + \ketbra{10}{10}_{12} \big) \ketbra{0}{1}_3\bigg) X_6}_{\text{maps } \ket{q_0q_1q_2q_3q_4q_5q_6} \text{ to } \ket{q_0 q_1 q_2 \overline{q_3} q_4 q_5 \overline{q_6}} \text{ if } q_1+q_2+q_3=0; \, \text{ returns } 0 \text{ otherwise}} \underbrace{\bigg. Z_1 Z_2 \bigg.}_{\mathclap{(-1)^{q_1+q_2}}}\\
 &=  \frac{1}{2}\big( \ketbra{00}{00}_{12} + \ketbra{11}{11}_{12} + \ketbra{01}{01}_{12} + \ketbra{10}{10}_{12} \big) X_3 X_6 Z_1 Z_2 \label{eqn:a3p1} \\ 
    & \quad  + \frac{i}{2} \bigg(\ketbra{01}{01}_{12} + \ketbra{10}{10}_{12} - \ketbra{00}{00} + \ketbra{11}{11}_{12} \bigg) Y_3 X_6 Z_1 Z_2  \nonumber \\
    &= \frac{1}{2}X_3 X_6 Z_1 Z_2 - \frac{i}{2} \bigg(\ketbra{01}{01}_{12} + \ketbra{10}{10}_{12} + \ketbra{00}{00} + \ketbra{11}{11}_{12} \bigg) Y_3 X_6 \label{eqn:a3p2} \\
    &= \frac{1}{2}\left(X_3 X_6 Z_1 Z_2 -i Y_3 X_6 \right) \, ,  \label{eqn:a3final}
\end{align}
using $\ketbra{1}{0} = \frac{1}{2}(X-iY)$ and $\ketbra{0}{1} = \frac{1}{2}(X+iY)$ to obtain Eq.~\ref{eqn:a3p1}, and absorbing the $Z_1Z_2$ operators into the bracketed quantity of the second term to produce the identity in Eq.~\ref{eqn:a3p2}. 

\end{document}

%% file: tikz_tree9.tex
\begin{figure}[h]
  \centering
  \scalebox{0.67}{
    \begin{tikzpicture}
      \draw
        (0, 0) node[circle,draw=black,fill=black!4,minimum size=1.05cm, label=center:\Large 0](0){}
        (1, 2) node[circle,draw=black,fill=black!4,minimum size=1.05cm, label=center:\Large 1](1){}
        (2, 0) node[circle,draw=black,fill=black!4,minimum size=1.05cm, label=center:\Large 2](2){}
        (2, 3) node[circle,draw=black,fill=black!4,minimum size=1.05cm, label=center:\Large 3](3){}
        (3, 5) node[circle,draw=black,fill=black!4,minimum size=1.05cm, label=center:\Large 4](4){}
        (4, 3) node[circle,draw=black,fill=black!4,minimum size=1.05cm, label=center:\Large 5](5){}
        (4, 0) node[circle,draw=black,fill=black!4,minimum size=1.05cm, label=center:\Large 6](6){}
        (5, 2) node[circle,draw=black,fill=black!4,minimum size=1.05cm, label=center:\Large 7](7){}
        (6, 0) node[circle,draw=black,fill=black!4,minimum size=1.05cm, label=center:\Large 8](8){};
      \begin{scope}[->]
        \draw (1) to (0);
        \draw (1) to (2);
        \draw (4) to[bend right](1);
        \draw (4) to[bend left](7);
        \draw (4) to (3);
        \draw (4) to (5);
        \draw (7) to (6);
        \draw (7) to (8);
      \end{scope}
    \end{tikzpicture}}
    \caption{The unpruned Sierpinski tree for $n = 9$}
    \label{fig:n9}
\end{figure}

%% file: tikz_tree27.tex
\begin{figure}[h]
  \centering
  \begin{tikzpicture}[scale=0.65]
      \draw
        (0, 0) node[circle,draw=black,fill=black!4,minimum size=0.7cm,label=center:0] (0){}
        (1, 2) node[circle,draw=black,fill=black!4,minimum size=0.7cm,label=center:1] (1){}
        (2, 0) node[circle,draw=black,fill=black!4,minimum size=0.7cm,label=center:2] (2){}
        (2, 3) node[circle,draw=black,fill=black!4,minimum size=0.7cm,label=center:3] (3){}
        (3, 5) node[circle,draw=black,fill=black!4,minimum size=0.7cm,label=center:4] (4){}
        (4, 3) node[circle,draw=black,fill=black!4,minimum size=0.7cm,label=center:5] (5){}
        (4, 0) node[circle,draw=black,fill=black!4,minimum size=0.7cm,label=center:6] (6){}
        (5, 2) node[circle,draw=black,fill=black!4,minimum size=0.7cm,label=center:7] (7){}
        (6, 0) node[circle,draw=black,fill=black!4,minimum size=0.7cm,label=center:8] (8){}
        (4, 7) node[circle,draw=black,fill=black!4,minimum size=0.7cm,label=center:9] (9){}
        (5, 9) node[circle,draw=black,fill=black!4,minimum size=0.7cm,label=center:10] (10){}
        (6, 7) node[circle,draw=black,fill=black!4,minimum size=0.7cm,label=center:11] (11){}
        (6, 10) node[circle,draw=black,fill=black!4,minimum size=0.7cm,label=center:12] (12){}
        (7, 12) node[circle,draw=black,fill=black!4,minimum size=0.7cm,label=center:13] (13){}
        (8, 10) node[circle,draw=black,fill=black!4,minimum size=0.7cm,label=center:14] (14){}
        (8, 7) node[circle,draw=black,fill=black!4,minimum size=0.7cm,label=center:15] (15){}
        (9, 9) node[circle,draw=black,fill=black!4,minimum size=0.7cm,label=center:16] (16){}
        (10, 7) node[circle,draw=black,fill=black!4,minimum size=0.7cm,label=center:17] (17){}
        (8, 0) node[circle,draw=black,fill=black!4,minimum size=0.7cm,label=center:18] (18){}
        (9, 2) node[circle,draw=black,fill=black!4,minimum size=0.7cm,label=center:19] (19){}
        (10, 0) node[circle,draw=black,fill=black!4,minimum size=0.7cm,label=center:20] (20){}
        (10, 3) node[circle,draw=black,fill=black!4,minimum size=0.7cm,label=center:21] (21){}
        (11, 5) node[circle,draw=black,fill=black!4,minimum size=0.7cm,label=center:22] (22){}
        (12, 3) node[circle,draw=black,fill=black!4,minimum size=0.7cm,label=center:23] (23){}
        (12, 0) node[circle,draw=black,fill=black!4,minimum size=0.7cm,label=center:24] (24){}
        (13, 2) node[circle,draw=black,fill=black!4,minimum size=0.7cm,label=center:25] (25){}
        (14, 0) node[circle,draw=black,fill=black!4,minimum size=0.7cm,label=center:26] (26){};
      \begin{scope}[->]
        \draw (1) to (0);
        \draw (1) to (2);
        \draw (4) to[bend right] (1);
        \draw (4) to[bend left] (7);
        \draw (4) to (3);
        \draw (4) to (5);
        \draw (7) to (6);
        \draw (7) to (8);
        \draw (10) to (9);
        \draw (10) to (11);
        \draw (13) to[out=197.5, in=90] (4);
        \draw (13) to[out=342.5, in=90] (22);
        \draw (13) to[bend right] (10);
        \draw (13) to[bend left] (16);
        \draw (13) to (12);
        \draw (13) to (14);
        \draw (16) to (15);
        \draw (16) to (17);
        \draw (19) to (18);
        \draw (19) to (20);
        \draw (22) to[bend right] (19);
        \draw (22) to[bend left] (25);
        \draw (22) to (21);
        \draw (22) to (23);
        \draw (25) to (24);
        \draw (25) to (26);
      \end{scope}
    \end{tikzpicture}
    \caption{The unpruned Sierpinski tree for $n = 27$}
    \label{fig:n27}
\end{figure}

%% file: mainv2.bbl
\begin{thebibliography}{10}

\bibitem{georgescu_quantum_2014}
I.M. Georgescu, S.~Ashhab, and Franco Nori.
\newblock Quantum simulation.
\newblock {\em Reviews of Modern Physics}, 86(1):153--185, March 2014.

\bibitem{kassal_simulating_2011}
Ivan Kassal, James~D. Whitfield, Alejandro Perdomo-Ortiz, Man-Hong Yung, and Alán Aspuru-Guzik.
\newblock Simulating {Chemistry} {Using} {Quantum} {Computers}.
\newblock {\em Annual Review of Physical Chemistry}, 62(1):185--207, May 2011.

\bibitem{Jordan:1928wi}
P.~Jordan and E.~Wigner.
\newblock About the {Pauli} exclusion principle.
\newblock {\em Zeitschrift für Physik}, 47(9-10):631--651, September 1928.

\bibitem{bravyi_fermionic_2002}
Sergey~B. Bravyi and Alexei~Yu. Kitaev.
\newblock Fermionic {Quantum} {Computation}.
\newblock {\em Annals of Physics}, 298(1):210--226, May 2002.

\bibitem{seeley_bravyi-kitaev_2012}
Jacob~T. Seeley, Martin~J. Richard, and Peter~J. Love.
\newblock The {Bravyi}-{Kitaev} transformation for quantum computation of electronic structure.
\newblock {\em The Journal of Chemical Physics}, 137(22):224109, December 2012.

\bibitem{havlicek_operator_2017}
Vojtěch Havlíček, Matthias Troyer, and James~D. Whitfield.
\newblock Operator locality in the quantum simulation of fermionic models.
\newblock {\em Physical Review A}, 95(3):032332, March 2017.

\bibitem{fenwick_new_1994}
Peter~M. Fenwick.
\newblock A new data structure for cumulative frequency tables.
\newblock {\em Software: Practice and Experience}, 24(3):327--336, March 1994.

\bibitem{fenwick_new_1996}
Peter~M. Fenwick.
\newblock A {New} {Data} {Structure} for {Cumulative} {Probability} {Tables}: {An} {Improved} {Frequency}-to-{Symbol} {Algorithm}.
\newblock {\em Software: Practice and Experience}, 26(4):489--490, April 1996.

\bibitem{jiang_optimal_2020}
Zhang Jiang, Amir Kalev, Wojciech Mruczkiewicz, and Hartmut Neven.
\newblock Optimal fermion-to-qubit mapping via ternary trees with applications to reduced quantum states learning.
\newblock {\em Quantum}, 4:276, June 2020.

\bibitem{gottesman_heisenberg_1998}
Daniel Gottesman.
\newblock The {Heisenberg} {Representation} of {Quantum} {Computers}.
\newblock July 1998.
\newblock arXiv: quant-ph/9807006.

\bibitem{obrien_local_2024}
Oliver O'Brien, Laurens Lootens, and Frank Verstraete.
\newblock Local {Jordan}-{Wigner} transformations on the torus, April 2024.
\newblock arXiv:2404.07727 [cond-mat, physics:math-ph, physics:quant-ph].

\bibitem{bultinck2017fermionic}
Nick Bultinck, Dominic~J. Williamson, Jutho Haegeman, and Frank Verstraete.
\newblock Fermionic matrix product states and one-dimensional topological phases.
\newblock {\em Phys. Rev. B}, 95:075108, Feb 2017.

\bibitem{steudtner2018fermion}
Mark Steudtner and Stephanie Wehner.
\newblock Fermion-to-qubit mappings with varying resource requirements for quantum simulation.
\newblock {\em New Journal of Physics}, 20(6):063010, jun 2018.

\bibitem{anticommuting2021sarkar}
Rahul Sarkar and Ewout Berg.
\newblock On sets of maximally commuting and anticommuting {Pauli} operators.
\newblock {\em Research in the Mathematical Sciences}, 8, 03 2021.

\bibitem{miller2023bonsai}
Aaron Miller, Zolt\'an Zimbor\'as, Stefan Knecht, Sabrina Maniscalco, and Guillermo Garc\'{\i}a-P\'erez.
\newblock Bonsai algorithm: Grow your own fermion-to-qubit mappings.
\newblock {\em PRX Quantum}, 4:030314, Aug 2023.

\bibitem{bataille_quantum_2020}
Marc Bataille.
\newblock Quantum circuits of {CNOT} gates, December 2020.
\newblock arXiv:2009.13247 [quant-ph].

\bibitem{harrison_sierpinski_2024}
Brent Harrison, Jason Necaise, Andrew Projansky, and James~D. Whitfield.
\newblock A {Sierpinski} {Triangle} {Data} {Structure} for {Efficient} {Array} {Value} {Update} and {Prefix} {Sum} {Calculation}, March 2024.
\newblock arXiv:2403.03990 [cs].

\bibitem{dehaene2003clifford}
Jeroen Dehaene and Bart De~Moor.
\newblock Clifford group, stabilizer states, and linear and quadratic operations over gf(2).
\newblock {\em Phys. Rev. A}, 68:042318, Oct 2003.

\bibitem{gunderman_transforming_2023}
Lane~G. Gunderman.
\newblock Transforming collections of {Pauli} operators into equivalent collections of {Pauli} operators over minimal registers.
\newblock {\em Physical Review A}, 107(6):062416, June 2023.

\bibitem{aaronson_improved_2004}
Scott Aaronson and Daniel Gottesman.
\newblock Improved simulation of stabilizer circuits.
\newblock {\em Physical Review A}, 70(5):052328, November 2004.

\end{thebibliography}
